\documentclass[twocolumn,showpacs,preprintnumbers,amssymb,prd]{revtex4}

\usepackage{graphicx,epsf, epsfig, amssymb}
\usepackage{bm}
\usepackage{longtable}

\usepackage{graphicx}
\usepackage{bm} 
\def\be{\begin{equation}}
\def\ee{\end{equation}}
\def\beq{\begin{eqnarray}}
\def\eeq{\end{eqnarray}}

\def\bes{\begin{eqnarray}}
\def\ees{\end{eqnarray}}

\newlength{\sizeonefig}
\newlength{\sizetwofig}
\begin{document}

\title{Geodesic stability, Lyapunov exponents and quasinormal modes}

\author{Vitor Cardoso} \email{vcardoso@fisica.ist.utl.pt}
\affiliation{Centro Multidisciplinar de Astrof\'{\i}sica -
CENTRA, Dept. de F\'{\i}sica, Instituto Superior T\'ecnico, Av. Rovisco
Pais 1, 1049-001 Lisboa, Portugal \&\\  Department of
Physics and Astronomy, The University of Mississippi, University,
MS 38677-1848, USA}

\author{Alex S. Miranda} \email{astmiranda@if.ufrj.br}
\affiliation{Instituto de F\'{\i}sica, Universidade Federal do Rio de Janeiro,
Caixa Postal 68528, RJ, 21941-972, Brazil}

\author{Emanuele Berti} \email{berti@wugrav.wustl.edu}
\affiliation{Jet Propulsion Laboratory, California Institute of Technology,
  Pasadena, CA 91109, USA\,\&\\  Department of
Physics and Astronomy, The University of Mississippi, University,
MS 38677-1848, USA}

\author{Helvi Witek} \email{helvi.witek@ist.utl.pt}
\affiliation{Centro Multidisciplinar de Astrof\'{\i}sica - CENTRA,
Dept. de F\'{\i}sica, Instituto Superior T\'ecnico, Av. Rovisco
Pais 1, 1049-001 Lisboa, Portugal
\& \\ Institute of Theoretical Physics, Friedrich-Schiller University Jena, Max-Wien-Platz 1, 07743 Jena, Germany}

\author{Vilson T. Zanchin} \email{zanchin@ufabc.edu.br}
\affiliation{Centro de Ci\^encias Naturais e Humanas,
Universidade Federal do ABC,Rua Santa Ad\'elia 166, 09210-170, Santo Andr\'e, SP, Brazil}


\begin{abstract}
  Geodesic motion determines important features of spacetimes. Null unstable
  geodesics are closely related to the appearance of compact objects to
  external observers and have been associated with the characteristic modes of
  black holes. By computing the Lyapunov exponent, which is the inverse of the
  instability timescale associated with this geodesic motion, we show that, in
  the eikonal limit, quasinormal modes of black holes in any dimensions are
  determined by the parameters of the circular null geodesics. This result is
  independent of the field equations and only assumes a stationary, spherically
  symmetric and asymptotically flat line element, but it does not seem to be
  easily extendable to anti-de Sitter spacetimes.  We further show that (i) in
  spacetime dimensions greater than four, equatorial circular timelike
  geodesics in a Myers-Perry black hole background are unstable, and (ii) the
  instability timescale of equatorial null geodesics in Myers-Perry spacetimes
  has a local minimum for spacetimes of dimension $d\geq6$.

\end{abstract}


\maketitle
\newpage
\section{Introduction}
Geodesics in black hole spacetimes have been extensively studied, both in four
and higher dimensional spacetimes, with and without a cosmological
constant. Geodesics may display a rich structure and they convey important
information on the background geometry. Among the different kinds of geodesic
motion, circular geodesics are especially interesting.  For instance, the
binding energy of the last stable circular timelike geodesic in the Kerr
geometry is related to the gravitational binding energy that can be radiated
to infinity, and it can be used to estimate the spin of astrophysical black
holes through observations of accretion disks
\cite{Narayan:2005ie,Zhang:1997dy,Shapiro:1983du}.

It was shown many years ago that null geodesics also play an important
role. The optical appearance of a star undergoing gravitational collapse
depends crucially on the circular unstable null geodesic, which also explains
an exponential fade-out of the collapsing star's luminosity
\cite{podurets,amesthorne}. Null geodesics are also very useful to explain the
characteristic modes of a black hole -- the so-called quasinormal modes (QNMs)
\cite{Kokkotas:1999bd,Nollert:1999ji}. These ``free'' modes of vibration can
be interpreted in terms of null particles trapped at the unstable circular
orbit and slowly leaking out
\cite{pressringdown,goebel,Ferrari:1984zz,mashhoon,Berti:2005eb}.  The real part of the
complex QNM frequencies is determined by the angular velocity at the unstable
null geodesic; the imaginary part is related to the instability timescale of
the orbit, a quantity which is seldom considered in geodesic studies, with
some noteworthy exceptions (see
e.g.~\cite{Bombelli:1991eg,Cornish:2003ig,Pretorius:2007jn,PerezGiz:2008yq,Levin:2008yp,Grossman:2008yk,Steklain:2008gz}).
Furthermore, there is some evidence \cite{Pretorius:2007jn} that unstable
circular orbits could yield information on phenomena occurring at the
threshold of black hole formation in the high-energy scattering of black
holes, a process of interest in fundamental physics for a variety of reasons
\cite{Sperhake:2008ga,Shibata:2008rq}.

In this work we clarify some aspects of the relation between unstable null
geodesics, Lyapunov exponents and quasinormal modes. In Section
\ref{LyapunovSec} we derive a simple formula for the principal Lyapunov
exponent $\lambda$ in terms of the second derivative of the effective
potential for radial motion $V_r$:
\be\label{LyapunovGeneralIntro}
\lambda=\sqrt{\frac{V_r''}{2 \dot{t}^2}}\,,
\ee
where $t$ is coordinate time. Throughout this work, a dot denotes a derivative
with respect to proper time and a prime stands for derivative with respect to
areal radius $r$. The result above is valid for a wide class of spacetimes and
geodesics, including stationary spherically symmetric spacetimes and equatorial
orbits in the geometry of higher-dimensional, rotating (Myers-Perry) black
hole solutions.

In Section \ref{sphericallysym} we show that the relation between QNMs and
unstable circular null geodesics is quite general, being valid in the eikonal
limit for any static, spherically symmetric, asymptotically flat
spacetime. More specifically, we show that the angular velocity $\Omega_c$ at
the unstable null geodesic and the Lyapunov exponent, determining the
instability timescale of the orbit (see for instance
\cite{Bombelli:1991eg,Cornish:2003ig}) agree with analytic WKB approximations
for QNMs \cite{Mashhoon:1983,Schutz:1985km,Iyer:1986np}:
\be\label{QNMLyapunovIntro}
\omega_{\rm QNM}=\Omega_c\,l-i(n+1/2)\,|\lambda|\,,
\ee
where $n$ is the overtone number and $l$ is the angular momentum of the
perturbation. The WKB results are formally valid only in the eikonal regime
($l\gg 1$), but they seem to yield surprisingly accurate predictions even for
low values of $l$ \cite{Iyer:1986nq,Berti:2005eb}. A simple derivation of the
Lyapunov exponent for spherically symmetric, asymptotically flat spacetimes,
patterned after the original QNM calculation by Ferrari and Mashhoon \cite{Ferrari:1984zz,mashhoon}, is
provided in Appendix \ref{sec:calcmashhoon}. For the important case of a
$d$-dimensional Schwarzschild-Tangherlini \cite{Tangherlini:1963bw} black hole
solution we find that the critical exponent defined by Pretorius and Khurana
\cite{Pretorius:2007jn} can be determined analytically to be
\be
\gamma\equiv
\frac{\Omega_c}{2\pi \lambda} = 
\frac{1}{2\pi\sqrt{d-3}}\,.
\label{tanghIntro}
\ee
By exploring the connection between QNMs and null geodesics, we also find a
simple analytical result for the quasinormal frequencies of near-extremal
Schwarzschild-de Sitter black holes in $d=4$:
\be\label{qnadsgeoIntro}
\omega_{\rm QNM}=\kappa_+\left[l-i\,(n+1/2)\right]\,,
\ee
where $\kappa_+$ denotes the surface gravity. In the eikonal limit, the above
result agrees with that found in \cite{Cardoso:2003sw}.

In Section \ref{formulation} we analyze the higher-dimensional
rotating black hole solutions found by Myers and Perry \cite{Myers:1986un}.
In $d=5$ we can compute $\lambda$ analytically. The Lyapunov exponent goes to zero as one
approaches extremality in $d=4,5$ spacetime dimensions. However, no such
behavior is observed for $d>5$: the Lyapunov exponent (normalized by the
orbital frequency) has a local minimum, which may be related to a possible
instability of the system first suggested by Emparan and Myers
\cite{Emparan:2003sy}. In Appendix \ref{sec:timelikemyers} we study in
some detail timelike circular geodesics in the Myers-Perry spacetime
and show that equatorial circular orbits are always unstable
for $d>4$. Finally, in Appendix \ref{SAdS} we discuss issues in
generalizing our results to non-asymptotically flat spacetimes.

\section{\label{LyapunovSec}Lyapunov exponents and geodesic stability}
Lyapunov exponents are a measure of the average rate at which nearby
trajectories converge or diverge in the phase space. A positive Lyapunov
exponent indicates a divergence between nearby trajectories, i.e., a high
sensitivity to initial conditions.  A geodesic stability analysis in terms of
Lyapunov exponents begins with the equations of motion schematically written
as 
\be \frac{d X_{i}}{dt} = H_{i}(X_{j}) \,, \ee
and linearized about a certain orbit:
\be \frac{d\,  \delta \! X_{i}(t)}{dt} = K_{ij}(t)\,  \delta \! X_{j}(t) \,. \ee
Here
\be K_{ij}(t) = \left. \frac{\partial H_{i}}{\partial X_{j}}\right|_{X_{i}(t)} \ee
is the linear stability matrix \cite{Cornish:2003ig}. The solution to the
linearized equations can be written as
\be \delta \! X_{i}(t) = L_{ij}(t)\, \delta\! X_{j}(0) \, \ee
in terms of the evolution matrix $L_{ij}(t)$, which must obey
\be \label{evo} \dot L_{ij}(t)=K_{im} L_{mj}(t)\, \ee
and $L_{ij}(0) = \delta_{ij}$. A determination of the eigenvalues of $L_{ij}$
leads to the principal Lyapunov exponent $\lambda$, which is the quantity we
focus on. Specifically
\be\label{L} \lambda = \lim_{t \rightarrow \infty} \frac{1}{t} \log \left( \frac{ L_{jj} (t)}{L_{jj} (0)}
\right) \, . \ee

We now restrict attention to a class of problems for which one has a two
dimensional phase space of the form $X_i(t)=(p_r, r)$. This includes circular
orbits in stationary spherically symmetric spacetimes and equatorial circular
orbits in stationary spacetimes, such as the Myers-Perry metric considered in
Section \ref{formulation}.  Linearizing the equations of motion with
$X_i(t)=(p_r, r)$ about orbits of constant $r$ we get
\be 
K_{ij} = \pmatrix{0 & K_1\cr K_2 & 0 } \,, 
\ee
where
\beq
K_1&=&\frac{d}{dr}\left (\dot{t}^{-1}\frac{\delta {\cal L}}{\delta r}\right )
\,,\\
K_2&=&-\left(\dot{t}\,g_{rr}\right)^{-1}\,,
\eeq
and ${\cal L}$ is the Lagrangian for geodesic motion (see below for explicit
examples).
Therefore, for circular orbits, the principal Lyapunov exponents can be
expressed as
\be
\lambda=\pm \sqrt{K_1\,K_2}\,.
\label{pL}\\
\ee
From now on we will drop the $\pm$ sign, and simply refer to the ``Lyapunov
exponent''. From the equations of motion it follows that
\be
\frac{d}{d\tau}\frac{\delta {\cal L}}{\delta \dot{r}}=\frac{\delta {\cal L}}{\delta r}\,,
\ee
and
\be
\frac{d}{d\tau}\frac{\delta {\cal L}}{\delta \dot{r}}=\frac{d}{d\tau}\left(-g_{rr} \dot{r}\right )=-\dot{r}\frac{d}{dr}(g_{rr}\dot{r})\,.
\ee
Using the definition of $V_r$, 
\be\label{Vrdefined}
\dot{r}^2=V_r\,,
\ee
we can rewrite this as
\be\label{dLdrgeneral}
\frac{\delta {\cal L}}{\delta r}=-\frac{1}{2g_{rr}}\frac{d}{dr}\left ( g_{rr}^2\,V_r\right)\,.
\ee
For circular geodesics $V_r=V_r'=0$ \cite{Bardeen:1972fi}, and Eq.~(\ref{pL})
reduces to
\be\label{LyapunovGeneral}
\lambda= 
\sqrt{\frac{V_r''}{2 \dot{t}^2}}\,.
\ee
Following Pretorius and Khurana \cite{Pretorius:2007jn}, we can define a
critical exponent
\be
\gamma \equiv \frac{\Omega_c}{2\pi \lambda} = \frac{T_\lambda}{T_\Omega}\,,
\ee
where we introduced a typical orbital timescale $T_\Omega\equiv 2\pi/\Omega_c$
and an instability timescale $T_{\lambda}\equiv 1/\lambda$ (note that in
Ref.~\cite{Cornish:2003ig} the authors use a different definition of the
orbital timescale, $T_{\Omega}\equiv 2\pi/\dot{\varphi}$, with $\varphi$ an
angular coordinate). Then we get
\be\label{Gamma}
\gamma= 
\frac{1}{2\pi}
\sqrt{\frac{\dot \varphi^2}{2 V_r''}}\,.
\ee
For circular null geodesics in many spacetimes of interest $V_r''>0$, which
implies instability. A quantitative characterization of this instability can
be achieved by computing the timescale associated with it.  In most of this
paper we will use the method outlined above (see also
\cite{Bombelli:1991eg,chaos,skokos}), but there are alternative approaches
\cite{Ferrari:1984zz,mashhoon,stewart,Cornish:2003ig}. In Appendix
\ref{sec:calcmashhoon}, for example, we recover the results of the next
Section following a stability analysis due to Ferrari and Mashhoon
\cite{Ferrari:1984zz,mashhoon}.

The discussion above is still rather general, assuming only that the variables
in the equations of motion form a two-dimensional plane in phase space. We now
specialize to spherically symmetric spacetimes.

\section{Spherically symmetric spacetimes \label{sphericallysym}}
We will consider a stationary spherically symmetric background
\be 
ds^{2}= f(r)dt^{2}- \frac{1}{g(r)} dr^{2}-r^{2}d\Omega_{d-2}^2\,,
\label{lineelementads} 
\ee
where $f(r)$ and $g(r)$ are functions to be determined by solving the field
equations, $d\Omega_{d-2}^2$ is the metric of the $(d-2)$-sphere and
$A_{d-2}\equiv 2 \pi^{(d-1)/2}/\Gamma{\left[(d-1)/2\right]}$ is the area of
the unit $(d-2)$-sphere. Since we do not specify the field equations, our
results are valid for any field equations admitting spherically symmetric,
asymptotically flat solutions.
The last property will be required to enforce the correct boundary
conditions in the WKB calculations of Section \ref{sec:WKBresults}.

\subsection{Circular orbits}
To compute the geodesics in the spacetime (\ref{lineelementads}) we follow
Chandrasekhar \cite{MTB}. Let us restrict attention to equatorial orbits, for
which the Lagrangian is
\be 2{\cal L}=f(r)\,\dot{t}^2-\frac{1}{g(r)}\dot{r}^2-r^2\dot{\varphi}^2 \,,
\label{lagrangianads}
\ee
where $\varphi$ is an angular coordinate.
The generalized momenta derived from this Lagrangian are
\beq p_t&=&f(r)\,\dot{t}\equiv E={\rm
const}\,,\label{pt2}\\
p_{\varphi}&=&-r^2\,\dot{\varphi}\equiv
-L={\rm const}\,,\label{pvarphi2}\\
p_r&=&-\frac{1}{g(r)}\dot{r}\,.\label{pr2}\eeq
The Lagrangian is independent of both $t$ and $\varphi$, so it follows
immediately that $p_t$ and $p_{\varphi}$ are two integrals of motion. Solving
(\ref{pt2})-(\ref{pvarphi2}) for $\dot{t}\,,\dot{\varphi}$ we get
\be
\dot{\varphi}=\frac{L}{r^2},\, \qquad \dot{t}=\frac{E}{f(r)}\,.\label{tdot}
\ee
The Hamiltonian is given by
\beq 2{\cal H}&=&2\left (p_t \dot{t}+p_{\varphi} \dot{\varphi}+p_r 
\dot{r}-{\cal L}\right )\nonumber \\
&=&f(r)\,\dot{t}^2-\frac{1}{g(r)}\dot{r}^2-r^2\,\dot{\varphi}^2 \nonumber\\
&=&E\dot{t}-L\dot{\varphi}-\frac{1}{g(r)}\dot{r}^2=\delta_1={\rm const}\,.
\label{2Ham}
\eeq
Here $\delta_1=1\,,0$ for time-like and null geodesics,
respectively. Inserting Eq.~(\ref{tdot}) in (\ref{2Ham}) and using 
the definition (\ref{Vrdefined}) we get
\be 
V_r=g(r)
\left[\frac{E^2}{f(r)}-\frac{L^2}{r^2}-\delta_1\right]\,.
\label{defVr}
\ee
%
\subsubsection{Timelike geodesics}
The requirement $V_r=V'_r=0$ for circular orbits yields
\be E^2= \frac{2f^2}{2f-r\,f'}\,,\quad
L^2=\frac{r^3\,f'}{2f-rf'}\,,\label{angmomentumcircular2} \ee
where here and below all quantities are evaluated at the radius of a circular
timelike orbit. Since the energy must be real, we require
\be
2f-rf'\,>\,0\,.\label{timelineconstraint}
\ee
The second derivative of the potential is
\be
V''_r=2\frac{g}{f}\frac{-3ff'/r+2(f')^2-ff''}{2f-rf'}\,,
\label{curvpotentialtimelike}
\ee
and the orbital angular velocity is given by
\be 
\Omega=
\frac{\dot{\varphi}}{\dot{t}}=
\left(\frac{f'}{2r}\right)^{1/2}\,. 
\ee
%
\subsubsection{Null geodesics}

Circular null geodesics satisfy the conditions:
\beq 
\frac{E}{L}&=&\pm \sqrt{\frac{f_c}{r_c^2}}\,,
\label{LElight}\\
2f_c&=&r_cf'_c\,,
\label{cirgeo} 
\eeq
Here and below a subscript $c$ means that the quantity in question is
evaluated at the radius $r=r_c$ of a circular null geodesic. An inspection of
(\ref{cirgeo}) shows that circular null geodesics can be seen as the innermost
circular timelike geodesics.  In this case
\be
V_r''(r_c)=
\frac{L^2g_c}{r_c^4f_c}
\left [2f_c-r_c^2f''_c\right]\,,
\label{V2p}
\ee
and the coordinate angular velocity is
\be
\Omega_c=
\frac{\dot{\varphi}}{\dot{t}}=
\left(\frac{f'_c}{2r_c}\right)^{1/2}=
\frac{f_c^{1/2}}{r_c}
\label{Omegacircular}\,.
\ee
%

\subsection{Lyapunov exponents}

\subsubsection{Timelike geodesics}

Using Eqs.~(\ref{LyapunovGeneral}), (\ref{tdot}) and
(\ref{curvpotentialtimelike})
to evaluate the Lyapunov exponent at
the circular timelike geodesics, we get
\beq\label{lamcirc2}
\lambda&=&\frac{1}{\sqrt{2}}\sqrt{-\frac{g}{f}
\left[\frac{3ff'}{r}-2(f')^2+ff''\right]} \, \nonumber\\
&=& \frac{1}{2}\sqrt{\left (2f-rf'\right )V''_r(r})\,.
\eeq
Bearing in mind that $2f-rf'>0$ and that unstable orbits are defined by
$V''_r>0$, we can see that $\lambda$ will be real whenever the orbit is
unstable, as expected.  In $d=4$ this formula reduces to well-known results
\cite{Cornish:2003ig}.  We also get

\be \frac{1}{\gamma^2}=
\left(\frac{2\pi \lambda}{\Omega}\right)^2=
(2\pi)^2\left[-
3g+2r\frac{g}{f}f'-r\frac{gf''}{f'}
\right]
\,.\ee

\subsubsection{Null geodesics}

Using Eqs.~(\ref{LyapunovGeneral}), (\ref{tdot}), (\ref{LElight}) and
(\ref{V2p}), for circular null geodesics we find
\be
\lambda= 
\frac{1}{\sqrt{2}}\sqrt{\frac{r_c^2f_c}{L^2}V_r''(r_c)}\,\nonumber\\
= \frac{1}{\sqrt{2}}\sqrt{-\frac{r_c^2}{f_c}\left(\frac{d^2}{dr_*^2}\frac{f}{r^2}\right)_{r=r_c}}\,.\label{lyaponov}
\ee
In the last equality we made use of (\ref{cirgeo}) and we defined the
``tortoise'' coordinate
\be
\frac{dr}{dr_*}=\sqrt{g(r)f(r)}\label{tortoise}\,.
\ee
%
\subsection{\label{sec:WKBresults} Unstable null geodesics and quasinormal modes: comparison with WKB results}

WKB methods \cite{Mashhoon:1983,Schutz:1985km,Iyer:1986np,Iyer:1986nq} provide
an accurate approximation of QNM frequencies in the eikonal limit for
spacetimes where the wave equation can be cast in the form
\be\label{WEWKB}
\frac{d^2}{dr_*^2}\Psi+Q_0\Psi=0\,,
\ee
where $r_*$ is a convenient ``tortoise'' coordinate, ranging from $-\infty$ to
$+\infty$. In particular, one gets the QNM condition
\be
\frac{Q_0(r_0)}{\sqrt{2Q_0^{(2)}(r_0)}}=i(n+1/2)\,,\label{wkb}
\ee
where $Q_0^{(2)}\equiv d^2Q_0/dr_*^2$ and Eq.~(\ref{wkb}) is evaluated at the
extremum of $Q_0$, i.e. the point $r_0$ at which $dQ_0/dr_*=0$.  We note that
this result is strictly valid only for asymptotically flat spacetimes, or for
spacetimes admitting wavelike solutions at spatial infinity.  It is not valid
for anti-de Sitter (AdS) backgrounds.  In a spacetime of the form
(\ref{lineelementads}), we find that the Klein-Gordon equation can be written
as in Eq.~(\ref{WEWKB}) with the tortoise coordinate (\ref{tortoise}). In the
eikonal limit ($l\rightarrow \infty$) we get
\be
Q_0\simeq \omega^2-f\frac{l^2}{r^2}\,.
\ee
It is known that scalar, electromagnetic and gravitational perturbations of
static black holes in higher dimensions have the same behavior in the eikonal
limit \cite{Kodama:2003jz,Ishibashi:2003ap,Kodama:2003kk}. In other words,
there is a well-defined geometric-optics (eikonal) limit where the potential
for a wide class of massless perturbations is ``universal''. For $Q_0$ above
we find that the extremum of $Q_0$ satisfies $2f(r_0)=r_0f'(r_0)$, i.e. $r_0$
coincides with the location of the null circular geodesic $r_0=r_c$, as given
by Eq.~(\ref{cirgeo}).  Furthermore, the WKB formula (\ref{wkb}) allows one to
conclude that, in the large-$l$ limit,
\be
\omega_{\rm QNM}=l\sqrt{\frac{f_c}{r_c^2}}
-i\frac{(n+1/2)}{\sqrt{2}}
\sqrt{-\frac{r_c^2}{f_c}\,\left (\frac{d^2}{dr_*^2}\frac{f}{r^2}\right )_{r=r_c}}\,.
\ee
Comparing with Eqs.~(\ref{Omegacircular}) and (\ref{lyaponov}) we find that
\be\label{QNMLyapunov}
\omega_{\rm QNM}=\Omega_c\,l-i(n+1/2)\,|\lambda|\,.
\ee
This is one of the main results of this paper: in the eikonal approximation,
the real and imaginary parts of the QNMs of any spherically symmetric,
asymptotically flat spacetime are given by (multiples of) the frequency and
instability timescale of the unstable circular null geodesics. 
\subsubsection{Higher-dimensional Schwarzschild  black hole}

Let us consider a more specific example: the higher-dimensional
Schwarzschild-Tangherlini metric \cite{Tangherlini:1963bw}
\be ds^{2}= fdt^{2}- f^{-1}dr^{2}-r^{2}d\Omega_{d-2}^2\,,\quad
f(r)=1-\left(\frac{r_+}{r}\right)^{d-3}\,,
\label{lineelementddim} 
\ee
which includes the well-known four-dimensional geometry as a special case.
Here $d\Omega_{d-2}^2$ is the metric of the $(d-2)$-sphere, and the horizon
radius $r_+$ is related to the mass $M$ of the spacetime via
$M=(d-2)A_{d-2}r_+^{d-3}/(16\pi)$.
For timelike geodesics we find that the orbits must satisfy
\be
r>r_c=\left (\frac{d-1}{2}\right )^{\frac{1}{d-3}}r_+\,,
\ee
where $r_c$ is the radius of the circular null geodesic.  With the requirement
(\ref{timelineconstraint}) we have $V''_r>0$ for all $d>4$, and therefore all
circular orbits are unstable for $d>4$ \cite{Tangherlini:1963bw,Rosa:2008dh}.
The four-dimensional case is special: one gets $V''_r=4M r
(6M-r)/(2r-6M)$. Thus, in four spacetime dimensions there are stable circular
orbits for any $r>6M$. The circular orbits with radius $3M<r<6M$ are all
unstable.  For light-like geodesics, one has $V_r''(r_c)=L^2\left (2d-6\right
)r_c^{-4}>0$. Therefore circular null geodesics are always unstable for $d\geq
4$.  The angular velocity at $r_c$ is given by
\be \Omega_c^2=\frac{d-3}{2}\,\left\lbrack
\frac{2}{d-1}\right\rbrack^{\frac{d-1}{d-3}}
\frac{1}{r_+^2}
\label{nullcircle}
\,.
\ee
The calculation of the Pretorius-Khurana \cite{Pretorius:2007jn} critical
exponent yields
\be
\gamma\equiv 
\frac{\Omega_c}{2\pi \lambda} = \frac{T_\lambda}{T_\Omega}=
\frac{1}{2\pi\sqrt{d-3}}\,,
\label{tangh}
\ee
where in the last equality we made use of Eq.~(\ref{nullcircle}). This result
is in excellent agreement with numerical calculations by Merrick and Pretorius
\cite{merrick}. Small values of $\gamma$ correspond to a strong Lyapunov
instability, so the instability is more pronounced for large spacetime
dimensions. By relating the geodesic orbital frequency and instability
timescale to the QNM frequencies in the eikonal limit, we get
\be
\frac{\omega_{\rm QNM}}{\Omega_c}=l-i\sqrt{d-3}\,(n+1/2)\,.
\ee
This is in complete agreement with known analytical and numerical results in
$d=4$ \cite{barreto} and higher dimensions
\cite{Konoplya:2003ii,Berti:2003si,Cardoso:2004cj}, for asymptotically flat
spacetimes.

\subsubsection{Near-extremal Schwarzschild-de Sitter spacetime in four dimensions}

A non-trivial example concerns a non-asymptotically flat spacetime, the
near-extremal Schwarzschild-de Sitter (SdS) spacetime in four
dimensions. General SdS spacetimes have a metric of the form
(\ref{lineelementads}) with $f(r)=g(r)=1 - 2M/r - (r/L_{\rm dS})^2$. $M$
denotes the black-hole mass and $L_{\rm dS}^2$ is the de Sitter curvature
radius, related to the cosmological constant $\Lambda$ by $L_{\rm dS}^2 =
3/\Lambda$.  The spacetime possesses two horizons: the black-hole horizon is
at $r=r_+$ and the cosmological horizon is at $r = r_{\rm Co}$, where $r_{\rm
  Co} > r_+$. The function $f$ has zeroes at $r_+$, $r_{\rm Co}$, and $r_0 =
-(r_+ + r_{\rm Co})$. In terms of these quantities, $f$ can be expressed as
\be
f = \frac{1}{L_{\rm dS}^2 r}\, (r-r_+)(r_{\rm Co}-r)(r-r_0).
\label{2.3}
\ee
It is useful to regard $r_+$ and $r_{\rm Co}$ as the two fundamental
parameters of the SdS spacetime, and to express $M$ and $L_{\rm dS}^2$ as
functions of these variables:
\beq
L_{\rm dS}^2 &=& {r_+}^2 + r_+ r_{\rm Co} + {r_{\rm Co}}^2\,,\\
\label{2.4}
2M L_{\rm dS}^2 &=& r_+ r_{\rm Co} (r_+ + r_{\rm Co})\,.
\label{2.5}
\eeq
We also introduce the surface gravity $\kappa_+$ associated with the black
hole horizon $r = r_+$: $\kappa_+ \equiv \frac{1}{2}
(df/dr)_{r=r_+}$. Explicitly, we have
\begin{equation}
\kappa_+ = \frac{(r_{\rm Co}-r_+)(r_+-r_0) }{ 2L_{\rm dS}^2 r_+ }.
\label{surface}
\end{equation}

Let us now specialize to the near-extremal SdS black hole, which is
defined as the spacetime for which the cosmological horizon $r_{\rm Co}$
is very close (in the $r$ coordinate) to the black hole horizon $r_+$, i.e.
\be
r_{\rm Co}-r_+\ll r_+\,.\label{defbhnariai}
\ee
For this spacetime one can make the following approximations:
\begin{equation}
r_0 \sim -2r_{+}^2\,\,;\,L_{\rm dS}^2 \sim 3r_{+}^2;\,\,
M \sim \frac{r_+}{3}\,\,;\,\kappa_+ \sim \frac{r_{\rm Co}-r_+}{2r_{+}^2}\,.
\label{approximation1}
\end{equation}
Note that $\kappa_+$ is infinitesimally small, since we assume Eq. (\ref{defbhnariai}) holds.
For null geodesics, an exact solution can be found with
\be r_c=\frac{3}{2} \left
(1-\frac{r_+^2}{L_{\rm dS}^2}\right
) \, r_+ \,.\label{nullcircleSdS}\ee
The angular velocity at this radius is given by
\be L_{\rm dS}^2\,\Omega_c^2=-1+
\frac{4}{27}
\left(1-\frac{r_+^2}{L_{\rm dS}^2}\right)^{-2}
\frac{L_{\rm dS}^2}{r_+^2}
\,,\label{circularSdS}\ee
and reduces to
\be
\Omega_c=\frac{(r_{\rm Co}-r_+)}{2r_+^2}=\kappa_+
\ee
in the near-extremal regime.  As in the case of four-dimensional Schwarzschild
black holes, Eq.~(\ref{tangh}), we find
\be 
\lambda/\Omega_c=1\,. 
\ee
This formula predicts
\be
\omega_{\rm QNM}=\kappa_+\left[l-i\,(n+1/2)\right]\,.\label{qnadsgeo}
\ee
The QNMs of this spacetime are known in closed form \cite{Cardoso:2003sw} and
agree with (\ref{qnadsgeo}) in the eikonal limit.

\section{\label{formulation}Myers-Perry black holes}
In four dimensions there is only one possible rotation axis for an
axisymmetric spacetime, and there is therefore only one angular momentum
parameter.  In higher dimensions there are several choices of rotation axis
and there is a multitude of angular momentum parameters, each referring to a
particular rotation plane. Rotating black hole solutions in higher dimensions
are known as Myers-Perry black holes \cite{Myers:1986un}.  We focus on the
simplest case for which there is only one angular momentum parameter, that we
shall denote by $a$.  The metric of a $d$-dimensional Myers-Perry black hole
with only one non-zero angular momentum parameter in Boyer-Lindquist-type
coordinates is given by (here we adopt the notation commonly used in related
works \cite{Ida:2002ez,Ida:2005ax,Ida:2006tf,Cardoso:2004cj,Cardoso:2005vk}),
\beq ds^2&=& {\Delta-a^2\sin^2\vartheta\over\Sigma}dt^2 
+{2a(r^2+a^2-\Delta)\sin^2\vartheta \over\Sigma}
dtd\varphi \nonumber\\
&&{}-{(r^2+a^2)^2-\Delta a^2 \sin^2\vartheta\over\Sigma}\sin^2\vartheta d\varphi^2
\nonumber\\
&&{}
-{\Sigma\over\Delta}dr^2
-{\Sigma}d\vartheta^2-r^2\cos^2\vartheta d\Omega_{d-4}^2,
\label{metric}
\eeq
where
\be
\Sigma=r^2+a^2\cos^2\vartheta\,,\quad \Delta=r^2+a^2-\mu r^{5-d}\,, 
\ee
and $d\Omega_{d-4}^2$ denotes the standard metric of the unit $(d-4)$-sphere
\cite{Myers:1986un}. This metric describes a rotating black hole in an
asymptotically flat, vacuum spacetime with mass and angular momentum
proportional to $\mu$ and $\mu a$, respectively. Hereafter we assume $\mu>0$
and $a>0$.

The event horizon is located at $r=r_+$ such that $\Delta|_{r=r_+}=0$. In the
standard four-dimensional case, an event horizon exists only for $a<\mu/2$. In
$d=5$ an event horizon exists only when $a<\sqrt{\mu}$, and it shrinks to zero
area in the extreme limit $a\rightarrow\sqrt{\mu}$.  On the other hand, when
$d\ge 6$, which is the part of the parameter space that we shall focus on,
$\Delta=0$ has exactly one positive root for arbitrary $a>0$.  This means
there is no bound on $a$, or (loosely speaking) that there are no extremal
Kerr black holes in higher dimensions.

\subsection{Circular geodesics in the equatorial plane}

To write down the geodesic equations in the Myers-Perry spacetime we follow
Chandrasekhar \cite{MTB}. Let us restrict attention to orbits in the
equatorial plane ($\dot{\vartheta}=0\,,\vartheta=\pi/2$), for which the
appropriate Lagrangian is
\be 2{\cal L}=g_{tt}\dot{t}^2 +2g_{t\varphi}\dot{t}\dot{\varphi}+
g_{rr}\dot{r}^2+g_{\varphi \varphi}\dot{\varphi}^2 \,. \label{lagrangian}
\ee
The generalized momenta following from this Lagrangian are
\beq p_t&=&g_{tt}\dot{t}+g_{t\varphi}\dot{\varphi}\equiv E={\rm
const}\,,\label{pt}\\
p_{\varphi}&=& g_{t\varphi}\dot{t}+g_{\varphi\varphi}\dot{\varphi}\equiv
-L={\rm const}\,,\label{pvarphi}\\
p_r&=& g_{rr}\dot{r}\,.\label{pr}\eeq
The Lagrangian is independent of both $t$ and $\varphi$, so it follows
immediately that $p_t$ and $p_{\varphi}$ are two integrals of motion. Solving
(\ref{pt})-(\ref{pvarphi}) for $\dot{t}\,,\dot{\varphi}$ we find
\beq \dot{\varphi}&=&\frac{1}{\Delta}\left [\frac{a\mu}{r^{d-3}}E
+\left(1-\frac{\mu}{r^{d-3}}\right )L\right
]\,,\label{dotvarphi}\\
\dot{t}&=&\frac{1}{\Delta}\left [\left(r^2+a^2+\frac{a^2\mu}
{r^{d-3}}\right )E-\frac{a\mu}{r^{d-3}}L\right
]\,.\label{dott}
\eeq
The Hamiltonian is given by
\beq 2{\cal H}&=&2\left (p_t \dot{t}+p_{\varphi} \dot{\varphi}+p_r \dot{r}
-{\cal L}\right )\,\nonumber \\
&=&\left (1-\frac{\mu}{r^{d-3}}\right )\dot{t}^2 
+\frac{2a\mu}{r^{d-3}}\dot{t}\dot{\varphi}-
\frac{r^2}{\Delta}\dot{r}^2\,\nonumber\\
&&-\left (r^2+a^2+\frac{a^2\mu}{r^{d-3}}\right )\dot{\varphi}^2 \,\nonumber\\
&=&E\dot{t}-L\dot{\varphi}-\frac{r^2}{\Delta}\dot{r}^2=\delta_1={\rm const}\,.\label{hamiltonian}\eeq
Here $\delta_1=1\,,0$ for time-like and null geodesics,
respectively. Inserting Eqs.~(\ref{dotvarphi})-(\ref{dott}) in Eq.~(\ref{hamiltonian})
we get
\beq\!\!\! \dot{r}^2&=&V_r\,,\label{potmyersperry} \\
\!\!\!r^2V_r&=&
\left [r^2E^2+\frac{\mu}{r^{d-3}}(aE-L)^2+(a^2E^2-L^2)
-\delta_1 \Delta \right]\,.\nonumber \eeq
In $d=4$, recalling that $\mu=2M$, we recover the well-known results for the
Kerr geometry \cite{MTB}. In five dimensions we recover the results by Frolov
and Stojkovic \cite{Frolov:2003en}, if we specialize their equations to only
one rotation parameter.

The conditions for the existence of circular geodesics, $V_{r}=V_{r}'=0$, lead
to the following equations:
\beq
0&=&r^{2}E^{2}+\mu r^{3-d}(aE-L)^2+(a^2 E^2-L^2)-
\delta_{1}\Delta\,,\nonumber\\
0&=&4r^{2}E^{2}-(d-5)\mu r^{3-d}(aE-L)^2+2(a^2 E^2-L^2)\nonumber\\
&&-\delta_{1}\left(2\Delta+r\Delta'\right)\,.
\label{condition2}
\eeq
Eliminating the term $a^2 E^2-L^2$ we get
\begin{equation}
2rE^2-(d-3)\mu r^{2-d}(aE-L)^2-\Delta'\delta_{1}=0.
\label{alternative1}
\end{equation}
%

\subsection{Circular null geodesics}

For light-like geodesics ($\delta_1=0$) we get the explicit conditions
\beq 0&=&r_c^2E^2+\mu r_c^{3-d}(aE-L)^2+(a^2E^2-L^2)\,,\label{null1}\\
0&=&2r_cE^2-(d-3)r_c^{2-d}\mu (aE-L)^2\,.\eeq
The above equations can be simplified by the introduction of the
impact parameter $D_c=L/E$:
\beq
0&=&r_{c}^{2}+\mu r_{c}^{3-d}(a-D_{c})^2+(a^2-D_{c}^2)\,,\label{null3}\\
0&=&2r_{c}-(d-3)\mu r_{c}^{2-d}(a-D_{c})^2\,.
\label{null4}
\eeq
From Eq.~(\ref{null4}) we get
\begin{equation}
D_{c}=a\mp\sqrt{\frac{2r_{c}^{d-1}}{(d-3)\mu}}.
\label{impactparameter} 
\end{equation}
Notice that Eq.~(\ref{null3}) is satisfied if and only if $|D_{c}|>a$. For
counterrotating orbits, we have $|D_{c}-a|=-(D_{c}-a)$ and this
case corresponds to the upper sign in the above equation, while for corotating
orbits, $|D_{c}-a|=+(D_{c}-a)$ and this case corresponds to the lower sign in
Eq. (\ref{impactparameter}).

Substituting Eq.~(\ref{impactparameter}) in (\ref{null3}), we find an equation
for the radius of circular null geodesics:
\begin{equation}
\frac{d-1}{d-3}r_{c}^2\pm 2a\sqrt{\frac{2r_{c}^{d-1}}{(d-3)\mu}}
-\frac{2r_{c}^{d-1}}{(d-3)\mu}=0.
\label{eqrc}
\end{equation}
In $d=4$ we recover the well-known results \cite{Bardeen:1972fi}.

An important quantity for the analysis of the null geodesics is the angular frequency at the null geodesic
$\Omega_{c}$:
\be
\Omega_{c}=
\frac{a\mu r_{c}^{3-d}+(1-\mu
r_{c}^{3-d})D_{c}}{(r_{c}^2+a^2+a^2\mu r_{c}^{3-d})-a\mu r_{c}^{3-d}D_{c}}
=\frac{1}{D_{c}}\,,
\ee
where we have used Eqs.~(\ref{impactparameter}) and (\ref{eqrc}).  Therefore
the frequency of equatorial null geodesics is the inverse of their impact
parameter. This generalizes the four-dimensional result \cite{MTB} to the
general case of Myers-Perry spacetimes. It would be interesting to investigate
whether or not this is a general property of any stationary spacetime.


For corotating orbits with $a\gg \mu^{1/(d-3)}$ and $d>5$, the following
analytical approximations are valid:
\beq
r_{\rm c}^{(d-5)/2}&\approx&
\frac{d-1}{2a\sqrt{2}}\sqrt{\frac{\mu}{d-3}}\,,
\nonumber
\\
D_c-a&\approx&\sqrt{\frac{2}{\mu(d-3)}}\left
  (\frac{d-1}{2a\sqrt{2}}\sqrt{\frac{\mu}{d-3}}\right )^{\frac{d-1}{d-5}}\,,
\nonumber
\\
&&({\rm corotating}\,,\,\,a\rightarrow \infty)\,.
\eeq
In particular, when the rotation is very large the radius of the corotating
orbit is at fixed relative distance from the horizon:
\be
\left (\frac{r_c^{\rm co}}{r_+}\right)^{\frac{d-5}{2}} \rightarrow \frac{d-1}{2\sqrt{2(d-3)}}\,.
\ee
This may be explained by the fact that the angular velocity of the horizon,
$\Omega_{r_+}=\frac{a}{r_+^2+a^2}$, also has a maximum and then decreases to
zero for very large $a$. 


On the other hand, counterrotating orbits are well described by
\beq
r_c^{(d-1)/2}&\approx& a\sqrt{2\mu(d-3)}\,,
\nonumber
\\
D_c&=&\frac{1}{\Omega_c}=-a\,,
\nonumber
\\
&&({\rm counterrotating}\,,\,\,a\rightarrow \infty)\,.
\eeq
We can express the radius of counterrotating orbits in terms of the horizon
radius in the limit of very large rotation:
\be
\left (\frac{r_c^{\rm counter}}{r_+}\right)^{\frac{d-1}{2}} \rightarrow \frac{a^2}{r_+^2}\sqrt{2(d-3)}\,.
\ee
Notice how counterrotating orbits {\it must} be located very far away from the
horizon as rotation increases. This is a consequence of having a ``strong''
ergoregion, extending through a large region in space.


For $d=5$ the previous equations simplify considerably, and we can find a
simple solution:
\beq\label{D5circ}
r_{c}&=&\sqrt{2}\sqrt{\mu\pm a\sqrt{\mu}}\,,\\
\frac{1}{D_c}&=&\Omega_c=\frac{1}{-a\mp 2\sqrt{\mu}}\,.
\eeq

\begin{figure*}[htb]
\begin{center}
\begin{tabular}{cc}
\epsfig{file=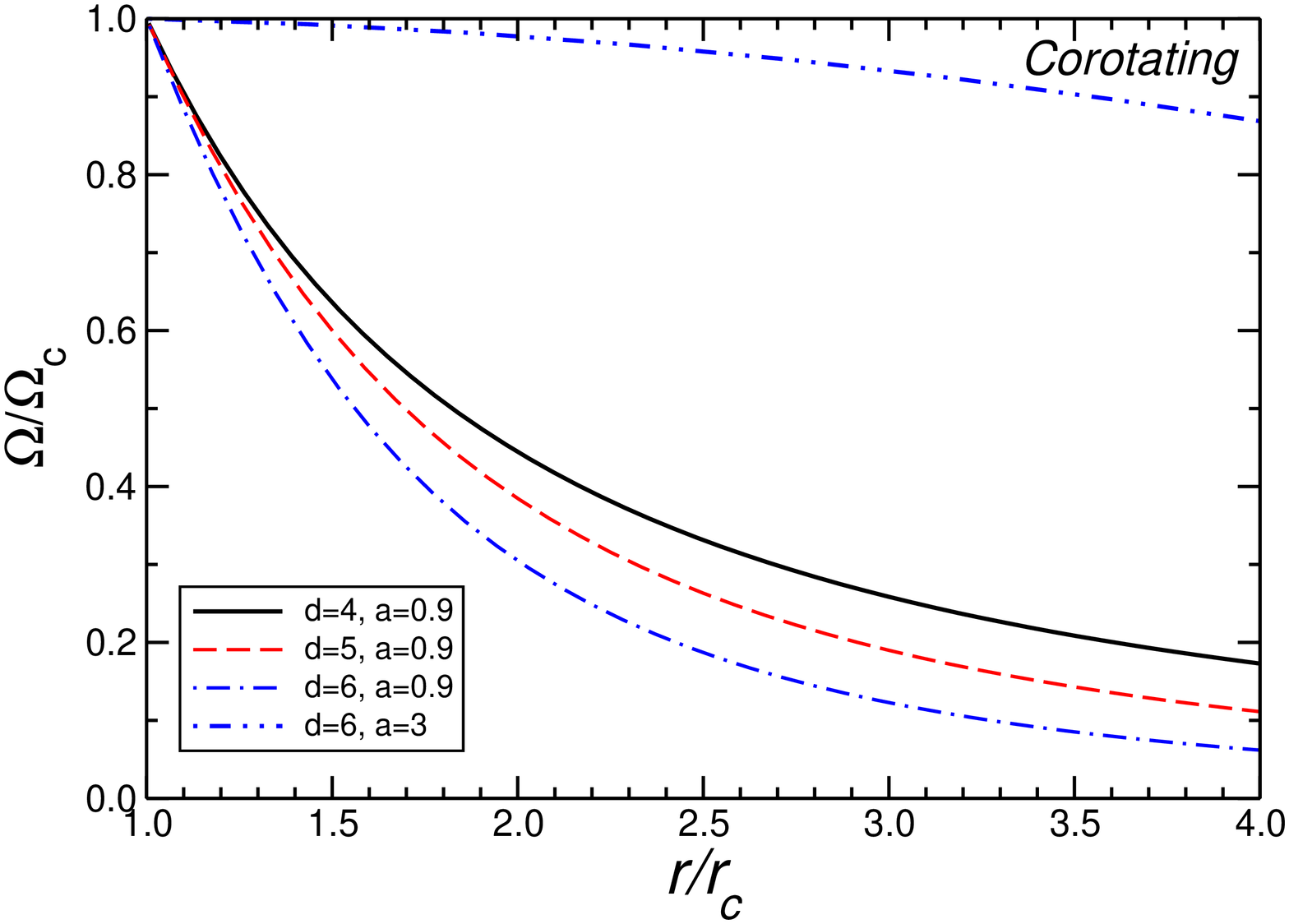,width=6cm,angle=270} &
\epsfig{file=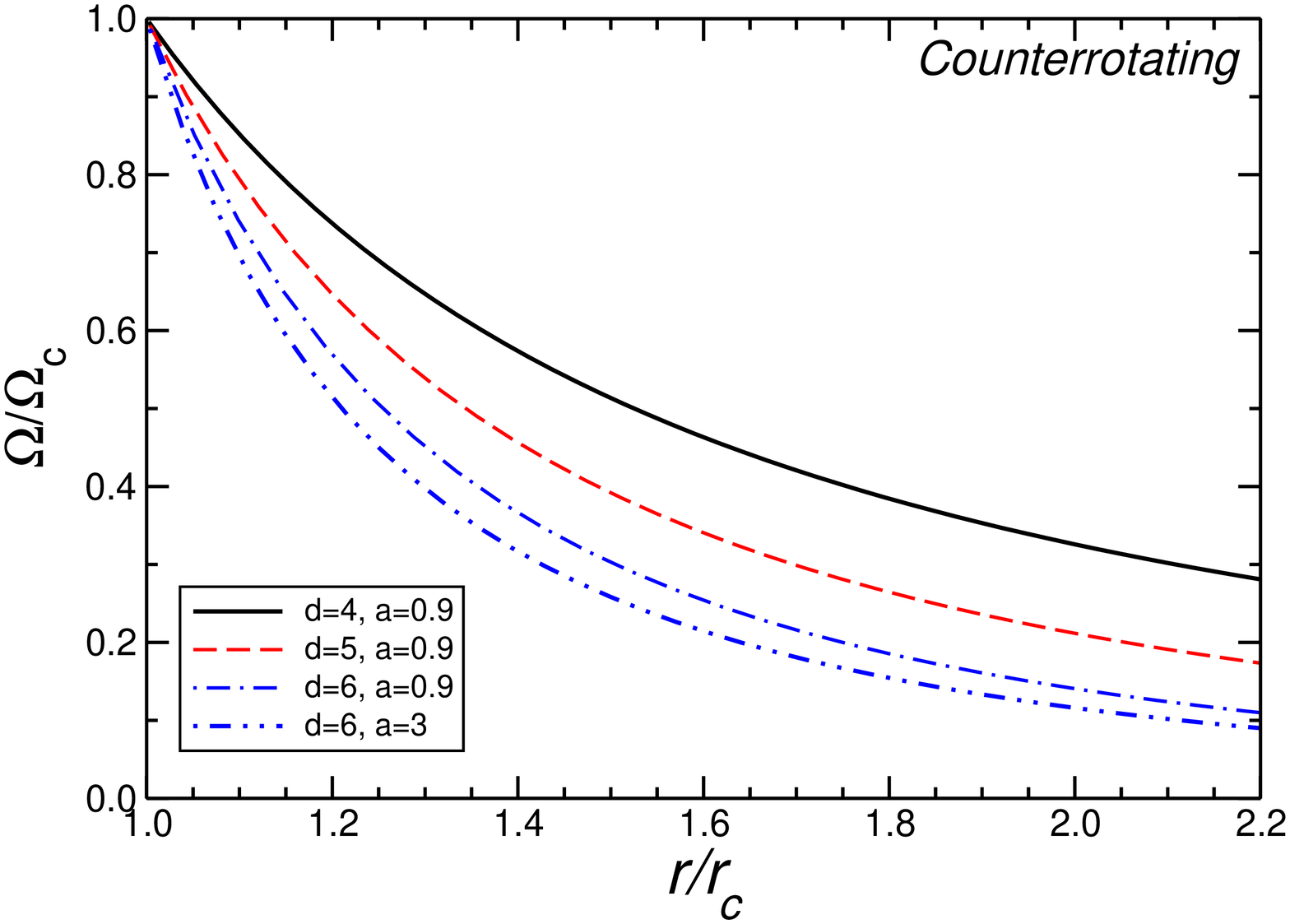,width=6cm,angle=270}
\end{tabular}
\caption{Ratio of the timelike orbital frequency $\Omega$ to the orbital
  frequency $\Omega_c$ of corotating (left) and counterrotating (right) null
  geodesics as a function of $r/r_c$, with $r_c$ the radius of the null
  geodesic.
  \label{fig:Timelike}}
\end{center}
\end{figure*}

\subsection{Circular timelike geodesics}

Timelike geodesics in a Myers-Perry spacetime are studied in Appendix
\ref{sec:timelikemyers}. In Section \ref{sec:instability}
we show that for $d>4$, there are no stable equatorial
circular orbits in this spacetime. This extends
the instability proof by Tangherlini \cite{Tangherlini:1963bw} to rotating
black holes, and the instability proof by Frolov and Stojkovic
\cite{Frolov:2003en} to a general number of spacetime dimensions.

The energy and angular momentum of timelike circular geodesics are studied
in Section \ref{sec:timelikegeo}, where we also consider the orbital frequency $\Omega$
of general circular timelike geodesics. In fact, it is possible to obtain a
simple expression for the ratio $\Omega/\Omega_{c}$ [see Eq.~(\ref{ratio1}) in Appendix \ref{sec:timelikemyers}]:
\be
\frac{\Omega}{\Omega_{c}}==\frac{\sqrt{2r_{c}^{d-1}}\mp
a\sqrt{(d-3)\mu}}{\sqrt{2r^{d-1}}\mp a\sqrt{(d-3)\mu}}\,.
\ee
This quantity is plotted in Fig.~\ref{fig:Timelike} for $\mu=2$ and selected
values of $a$ and $d$.  There is clearly a change in behavior for $d>5$ and
large rotation, exemplified here for $d=6$. This will be explored in more
detail in the next Section.

\subsection{Lyapunov exponents}

Applying Eq.~(\ref{LyapunovGeneral}) to the case of Myers-Perry black holes we
find
\be
\frac{\lambda}{\Omega_c}=\frac{
\sqrt{d-1}\left (-\mu\,r_c^4+a^2r_c^{d-1}+r_c^{d+1}\right)
}{
a r_c^d+\sqrt{\frac{2r_c^{d-1}}{\mu(d-3)}}(r_c^d-\mu\, r_c^3)
}\,.
\ee

\begin{figure}[t]
\begin{center}
\begin{tabular}{c}
\epsfig{file=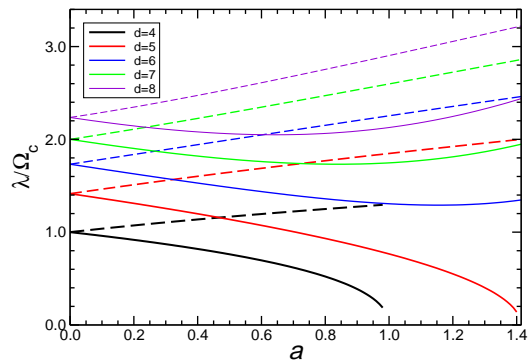,width=6cm,angle=270} 
\end{tabular}
\caption{Dimensionless instability exponents $\lambda/\Omega_c$ as a function
  of rotation for several spacetime dimensions $d$. We use units such that
  $\mu=2$. Solid lines refer to corotating orbits, dashed lines to
  counterrotating orbits.
  \label{fig:Lyapunov}}
\end{center}
\end{figure}

Using Eq.~(\ref{D5circ}) we can analytically compute the $d=5$ case, and we
get
\beq
\lambda^{\rm counter}\sqrt{\mu}&=&\sqrt{2}\frac{\sqrt{\mu-\sqrt{\mu}\,a}}{a-2\sqrt{\mu}}\,,\\
\lambda^{\rm co}\sqrt{\mu}&=&\sqrt{2}\frac{\sqrt{\mu+\sqrt{\mu}\,a}}{a+2\sqrt{\mu}}\,.
\eeq
The general case requires a numerical treatment. In Fig.~\ref{fig:Lyapunov} we
show the Lyapunov exponent normalized by the angular velocity,
$\lambda/\Omega_c$, as a function of rotation $a$.

For corotating geodesics in $d=4$ and $d=5$, $\lambda$ asymptotes to zero near
extremality ($a\to \mu/2$ and $a\to \sqrt{\mu}$, respectively). This does not
happen for $d>5$ and large rotation parameters: in this case,
$\lambda/\Omega_c$ has a local minimum. This is consistent with QNM
calculations in five \cite{Berti:2003yr} and higher dimensions
\cite{Cardoso:2004cj}. The local minimum may be related to a change in
behavior corresponding to a black hole-$\rightarrow$ black brane transition,
observed in \cite{Emparan:2003sy} in relation with a conjectured instability
of these systems for very large rotation rates.
\vskip 1mm
\begin{table}
  \caption{\label{tab:transition} The transition point $a_{\rm trans}$. The 
  second column refers to the critical value of rotation $a$ for which the
  corotating Lyapunov exponent has a minimum. The third column refers to the
  point at which the temperature has a minimum, an indicator considered in
  \cite{Emparan:2003sy}.
  }
\begin{tabular}{ccc}  \hline
\multicolumn{1}{c}{} 
$d$ & {\rm Lyapunov}               &     {\rm Temperature}          \\ 
\hline
\hline
6   &  1.15  
&  1.37 \\ 
7   &  0.84
&  1.28 \\ 
8   &  0.64   
&  1.22  \\ 
50  &  0.05   
&  1.02  \\ 
\end{tabular}
\end{table}
\vskip 1mm
The locations of minima in $\lambda/\Omega_c$ are given in Table
\ref{tab:transition}.  In this Table we also list a quantity considered in
\cite{Emparan:2003sy} as a possible indicator of a ``transition'' point, in
this case the rotation at which the temperature has a minimum. The two
quantities are roughly consistent for small $d$, but not for large $d$.

\section{Conclusions and Future Work}

We have shown that for all spherically symmetric spacetimes, in a geometrical
optics approximation, QNMs can be interpreted as particles trapped at unstable
circular null geodesics and slowly leaking out.  The leaking timescale is
given by the principal Lyapunov exponent, for which we obtained a fairly
simple expression, Eq.~(\ref{LyapunovGeneral}), in terms of the second
derivative of the effective radial potential for geodesic motion. This simple,
intuitive relation between QNMs and circular null geodesics is valid for all
asymptotically flat, spherically symmetric black hole spacetimes. 

Some aspects of our investigation deserve further analysis. The interpretation
of QNMs in terms of unstable circular null geodesics is valid in all
generality only for spherically symmetric, asymptotically flat spacetimes.
Once we break the azimuthal degeneracy, things get a bit more complex. For
instance, in $d=4$ it is known that equatorial geodesics can account for the
$l=|m|$ modes of Kerr-Newman black holes
\cite{Ferrari:1984zz,Berti:2005eb,Berti:2005ys}, but it is unclear whether this
analogy can be extended to modes with $l\neq |m|$.  Perhaps modes with $l \neq
m$ can be explained in terms of more general (e.g., non-equatorial)
geodesics. Besides improving our intuitive physical understanding of ringdown
radiation, a deeper exploration of this analogy could have important
implications for the interpretation of numerical simulations of black hole
binary mergers and their use in gravitational-wave data analysis
\cite{Berti:2007fi,Hanna:2008um,Baker:2008mj}.

Another limitation of our results concerns their extension to non
asymptotically flat (e.g., AdS) backgrounds.  In the large damping limit, a
certain class of QNMs has been associated with {\it radial} geodesics
\cite{Fidkowski:2003nf,Amado:2008hw,Festuccia:2008zx}. It would be very
interesting to extend this analysis to the eikonal (large-$l$) limit. A
possible starting point could be the $2+1$ dimensional
Ba\~nados-Teitelboim-Zanelli (BTZ) black hole \cite{Banados:1992wn}, for which
QNM frequencies are known analytically \cite{Cardoso:2001hn}. Quite apart from
the geodesic analogy, the large-$l$ limit is interesting {\it per se}.  It
turns out that the imaginary part of Schwarzschild-anti-de Sitter (SAdS) QNMs
decreases with $l$ \cite{Festuccia:2008zx}. Thus, if excited considerably,
large-$l$ modes could dominate the black hole's response to perturbations. A
more thorough investigation of the large-$l$ limit of QNMs is necessary.

Interesting physical phenomena could occur in (hypothetical) spacetimes for
which timelike circular geodesics have a frequency equal to (or larger than)
the frequency of unstable null geodesics. This would raise the interesting
possibility of exciting QNMs by orbiting particles, possibly leading to
instabilities of the spacetime.  It would be interesting to find general
conditions under which spacetimes possess stable null geodesics; stable null
geodesics may also be associated with instability (or marginal stability) of
spacetimes.

\section*{Acknowledgements}

We are grateful to Christopher Merrick and Frans Pretorius for sharing
unpublished results with us. We thank Marc Casals, Curt Cutler, Paolo Pani and Jorge Rocha 
for a critical reading of the manuscript and useful suggestions.
V.C.'s work was partially funded by Funda\c c\~ao
para a Ci\^encia e Tecnologia (FCT) - Portugal through projects
PTDC/FIS/64175/2006 and POCI/FP/81915/2007 and by a Fulbright Scholarship.
A.S.M. thanks the CENTRA/IST for hospitality and financial help and the
Conselho Nacional de Desenvolvimento Cient\'\i fico e Tecnol\'ogico (CNPq),
Brazil, for a grant.
H.W.'s work was partly supported by DFG grant SFB/ Transregio 7, Germany, and by Funda\c c\~ao para a Ci\^encia e Tecnologia (FCT) -
Portugal through grant SFRH/BD/46061/2008 and project PTDC/FIS/64175/2006.
V.T.Z. is partially supported by a fellowship from Conselho Nacional de
Desenvolvimento Cient\'\i fico e Tecnol\'ogico (CNPq) - Brazil.
E.B.'s research was supported by the NASA Postdoctoral Program at JPL,
administered by Oak Ridge Associated Universities through a contract with
NASA.
%


\appendix

\section{\label{sec:calcmashhoon} The instability timescale of circular geodesics: a
simpler derivation}

Perhaps the simplest way to determine the instability timescale associated
with circular null geodesics is through a consideration of the equations
defining these geodesics \cite{amesthorne,Ferrari:1984zz,mashhoon}. Indeed,
this was the approach originally adopted by Ferrari and Mashhoon to compute
quasinormal modes of Kerr-Newman black holes in the eikonal limit
\cite{Ferrari:1984zz,mashhoon}. In this Appendix we will rederive the
instability parameter $\lambda$ within their approach.

Consider small perturbations of a bundle of test null rays in the unstable
equatorial circular orbit around a black hole described by the metric
(\ref{lineelementads}).  First, rescale the affine parameter $s$ so as to be
the coordinate time $t$, and consider the following values of the unperturbed
geodesic:
\begin{equation}
t = s\,,\,\,\, r = r_c\,,\,\,\, \theta = \frac{\pi}{2} \,,\,\,\, \phi = \Omega_c\,s\,.
\end{equation}
The slightly perturbed equatorial null orbit is given by
\begin{eqnarray}
s      & = & t + \epsilon b(t) \,,\\
r      & = & r_c (1 + \epsilon h(t)) \,,\\
\theta & = & \pi/2 \,,\\
\phi   & = & \Omega_c (t + \epsilon k(t))\,,
\end{eqnarray}
where $|\epsilon| \ll 1$ denotes the dimensionless amplitude of the
perturbation.  Considering the leading terms in
Eqs. (\ref{tdot})-(\ref{defVr}), together with the boundary conditions that
$b(t)$, $h(t)$ and $k(t)$ vanish at $t = 0$, yields
\begin{eqnarray}
h(t) & = & \sinh(\lambda t) \,,\\
b(t) & = & \frac{r_c f'_c}{E \lambda}(\cosh(\lambda t) -1) \,,\\
k(t) & = & 0  \,.
\end{eqnarray}
The parameter $\lambda$ characterizes the decay rate and is determined by
\begin{equation}
\lambda^2 = \frac{f_c^2}{2E^2}\,V_r''(r_c)=\frac{r_c^2\,f_c}{2L^2}\,V_r''(r_c)\,,
\end{equation}
This result agrees with Eq.~(\ref{lyaponov}), obtained through the principal
Lyapunov exponent.

Finally, the instability timescale can also be derived (at least in the usual
Schwarzschild geometry) by considering a special class of geodesics:
inspiralling geodesics that asymptote to the light ring when $t\rightarrow
\infty$. These geodesics are considered in Chandrasekhar's book \cite{MTB},
and it is straightforward to compute how they approach the light ring.

\section{\label{sec:timelikemyers} Timelike geodesics in the equatorial plane of Myers-Perry spacetimes}

In a $d$-dimensional Myers-Perry black hole with only one non-zero angular
momentum parameter, the radial equation for geodesics in the equatorial plane
can be cast in the form $r^{4}\dot{r}^{2}=V$ with $V\equiv r^4V_r$ as given in
Eq. (\ref{potmyersperry}). The first derivative of $V$ with respect to $r$ is
given by
\beq
V'&=&4r^{3}E^{2}-(d-5)\mu r^{4-d} (aE-L)^2+2r(a^2 E^2-L^2)\nonumber\\
&&-\delta_{1}\left[4r^{3}+2ra^{2}+(d-7)\mu r^{6-d}\right]\,.
\label{firstderivative}
\eeq
The conditions for the existence of circular orbits are $V=0$ and $V'=0$:
\beq
0&=&r^{2}E^{2}+\mu r^{3-d}(aE-L)^2\nonumber\\
&+&(a^2 E^2-L^2)-\delta_{1}\Delta\,,\label{condition133}\\
0&=&4r^{2}E^{2}-(d-5)\mu r^{3-d}(aE-L)^2+2(a^2 E^2-L^2)\nonumber \\
&-&\delta_{1}\left(2\Delta+r\Delta'\right)\,.
\label{condition233}
\eeq
Eliminating the term $a^2 E^2-L^2$ one finds
\begin{equation}
2rE^2-(d-3)\mu r^{2-d}(aE-L)^2-\Delta'\delta_{1}=0\,.
\label{alternative122}
\end{equation}

\subsection{\label{sec:instability} The instability of geodesics in the
equatorial plane of Myers-Perry spacetimes}

The equations (\ref{condition133}) and (\ref{alternative122}) will be used
here to obtain the values of energy $E$ and angular momentum $L$ associated to
circular timelike ($\delta_{1}=1$) geodesics. Using (\ref{alternative122}) and
introducing the new quantities $x=L-aE$, $M=\mu/2$, and the reciprocal radius
$u=1/r$, we obtain an expression for $E^{2}$:
\begin{equation}
E^{2}=\left[1-(5-d)Mu^{d-3}\right]+(d-3)Mu^{d-1}x^2\,.
\label{en1app}
\end{equation}
With this expression for $E^{2}$, Eq.~(\ref{condition133}) leads to
\begin{eqnarray}
2aExu& =& x^{2}\left[(d-1)Mu^{d-3}-1\right]u \nonumber\\
& &-\left[a^{2}u-(d-3)Mu^{d-4}\right]\,.
\label{en2app}
\end{eqnarray}
For $d=4$ spacetime dimensions, the above equations are identical to those
obtained by Chandrasekhar \cite{MTB}.  We can now eliminate $E$ in equations
(\ref{en1app}) and (\ref{en2app}) to obtain a quadratic equation for $x^2$:
\beq
\label{eqforx}
\hskip -1cm  0&\!=\!&\!x^4 u^2 \left[\left[(d-1)Mu^{d-3}-1\right]^{2}-4a^2 M(d-3)u^{d-1}\right]\nonumber\\
&&\!\!\!\!-2x^2 u{\biggl[} \left[(d-1)Mu^{d-3}-1\right]\times\left[a^2 u-(d-3)Mu^{d-4}\right]\nonumber\\
&&\!\!\!\!-2a^2 u \left[(5-d)Mu^{d-3}-1\right]{\biggr]}\nonumber\\
&& +\left[a^2 u-(d-3)Mu^{d-4} \right]^2\,.
\eeq
The discriminant associated to this equation is given by $16Ma^2
(d-3)u^{d-1}\Delta_{u}^2$, where we have introduced $\Delta_{u}=a^2
u^2-2Mu^{d-3}+1$. In order to write the solutions of equation (\ref{eqforx}),
it is convenient to consider the expression
\begin{equation}
\left[(d-1)M u^{d-3}-1\right]^{2}-4a^2 M (d-3)u^{d-1}=Q_{-}Q_{+}\,,
\end{equation}
where
\begin{equation}
Q_{\pm}=1-(d-1)Mu^{d-3}\pm 2a\sqrt{(d-3)Mu^{d-1}}\,.
\end{equation}
Then the solutions of Eq.~(\ref{eqforx}) can be written as
\begin{equation}
x^2 u^2=\frac{Q_{\pm}\Delta_{u}-Q_{+}Q_{-}}{Q_{+}Q_{-}}=
\frac{1}{Q_{\mp}}(\Delta_{u}-Q_{\mp})\,.
\end{equation}
As an alternative, we can use the identity
\begin{equation}
\Delta_{u}-Q_{\mp}=u\left[a\sqrt{u}\pm\sqrt{(d-3)Mu^{d-4}}\right]^2\,,
\end{equation}
to cast the solution for $x$ in the simple form
\begin{equation}
x=-\frac{a\sqrt{u}\pm\sqrt{(d-3)Mu^{d-4}}}{\sqrt{uQ_{\mp}}}\,,
\label{expressionforx}
\end{equation}
where the upper sign in the foregoing equations applies to counterrotating
orbits, while the lower sign applies to corotating orbits.  Inserting
expression (\ref{expressionforx}) in equation (\ref{en1app}) and using the relation $L=aE+x$, we get the following expression for $E$,
\begin{equation}
E=\frac{1}{\sqrt{Q_{\mp}}}\left[1-2Mu^{d-3}\mp a\sqrt{(d-3)Mu^{d-1}}\right]\,.
\label{efinal}
\end{equation}
and the angular momentum associated with the circular geodesics,
\begin{equation}
L=\mp\frac{\sqrt{(d-3)Mu^{d-4}}}{\sqrt{uQ_{\mp}}}\left[1+a^2 u^2
\pm 2a\sqrt{\frac{Mu^{d-1}}{d-3}}\right]\,.
\label{lfinal}
\end{equation}

In order to investigate the stability of circular timelike orbits we must
compute the second derivative of $V$ with respect to $r$ for the values of $E$
and $L$ specific to circular orbits. Differentiating
Eq.~(\ref{firstderivative}) we find
\beq
 V''&\! =&\! 12r^2(E^2-1)+2(d-4)(d-5)Mr^{3-d}x^2   -   2x^2 \nonumber\\
& &\!\! -4aEx- 2a^2+ 2(d-6)(d-7)Mr^{5-d}\,.
\label{secondderivative1}
\eeq
Substituting $x$ and $E$ from Eqs.~(\ref{expressionforx}) and (\ref{efinal}),
the above expression for $V''$ becomes
\beq
V''&=&\frac{2(d-3)M u^{d-5}}{Q_{\mp}}{\biggl[}2(d-1)Mu^{d-3}+(d-5)\nonumber\\
&&\pm 8a\sqrt{(d-3)Mu^{d-1}}+(d-1)a^2 u^2{\biggr]}\,.
\label{secondderivative2}
\eeq
The term within square brackets in the foregoing equation is equal to
$(d-1)\Delta_{u}-4Q_{\mp}$, so that the second derivative of $V$ reduces to
\begin{equation}
V''=\frac{2(d-3)M u^{d-5}}{Q_{\mp}}\left[(d-1)\Delta_{u}-4Q_{\mp}\right]\,.
\label{secondderivative3}
\end{equation}
This expression shows an explicit dependence on the spacetime dimensionality
$d$.  To analyze the sign of $V''$ in Eq.~(\ref{secondderivative3}) it will be
helpful to distinguish between different values of $d$. Since $E$, $L$ and
$x=L-aE$ must be real, the functions $\Delta_{u}$ and $Q_{\pm}$ are such that
\begin{equation}
\Delta_{u}\geq Q_{\pm}\geq 0\,.
\end{equation}
For $d\geq 5$ the above conditions lead to
\begin{equation}
(d-1)\Delta_{u}\geq 4Q_{\mp}\qquad\Longrightarrow\qquad V''\geq 0\,.
\end{equation}
This means that there are no stable timelike circular orbits for spacetimes
with $d\geq 5$. This generalizes previous work by Tangherlini on non-rotating
higher-dimensional black holes \cite{Tangherlini:1963bw} and by Frolov and
Stojkovic on five-dimensional rotating black holes \cite{Frolov:2003en}.

\subsection{\label{sec:timelikegeo} The orbital frequency of circular geodesics}

The orbital frequency $\Omega=d\varphi/dt$ associated to circular timelike
geodesics is given by
\be
\Omega=
\frac{\left(L-2Mu^{d-3}x\right)u^2}{\left(1+a^2u^2\right)E-2aMu^{d-1}x}\,.
\label{freqtimelike12}
\ee
The foregoing expression can be simplified by considering the following
identities:
\beq
&&L-2Mu^{d-3}x=\mp\frac{\sqrt{(d-3)Mu^{d-4}}}{\sqrt{uQ_{\mp}}}\Delta_{u}\,,
\label{identity1}\\
&&\left(1+a^2u^2\right)E-2aMu^{d-1}x\nonumber \\
&&=\frac{\Delta_{u}}{\sqrt{Q_{\mp}}}\left[
1\mp a\sqrt{(d-3)Mu^{d-1}}\right]\,.\label{identity2}
\eeq
Substituting (\ref{identity1}) and (\ref{identity2}) into equation
(\ref{freqtimelike12}), we obtain
\begin{equation}
\Omega=\frac{\mp\sqrt{(d-3)Mu^{d-1}}}{1\mp a\sqrt{(d-3)Mu^{d-1}}}\,.
\label{freqtimelike2}
\end{equation}

%
%
By considering equation (\ref{impactparameter}) for $D_c$ and the relation
$\Omega_c=1/D_c$, we find a similar expression for circular null geodesics:
\begin{equation}
\Omega_{c}=\frac{\mp\sqrt{(d-3)Mu_{c}^{d-1}}}{1\mp
a\sqrt{(d-3)Mu_{c}^{d-1}}}\,.
\end{equation}
Consequently, the ratio $\Omega/\Omega_{c}$ varies with $r$ as follows:
\begin{equation}
\frac{\Omega}{\Omega_{c}}=\frac{\sqrt{r_{c}^{d-1}}\mp
a\sqrt{(d-3)M}}{\sqrt{r^{d-1}}\mp a\sqrt{(d-3)M}}\,.
\label{ratio1}
\end{equation}

\section{\label{SAdS} Schwarzschild-anti-de Sitter spacetimes}

In this Appendix we consider Schwarzschild-anti-de Sitter spacetimes, and we
show that the analogy between unstable circular orbits and black hole QNMs is
not trivially extended to non-asymptotically flat backgrounds. A
higher-dimensional Schwarzschild-anti-de Sitter solution is a solution of
\be
G_{ab}+\frac{(d-2)(d-1)}{2L_{\rm ads}} g_{ab}=0\,,
\ee
with cosmological constant $\Lambda\equiv -(d-2)(d-1)/(2L_{\rm ads})$
and typical curvature radius $L_{\rm ads}$. We consider the simplest black
hole solution: the $d$-dimensional Schwarzschild-anti-de Sitter solution. 
The line element is given by Eq.~(\ref{lineelementads}) with
\be
f(r)=g(r)=\left(\frac{r^{2}}{L_{\rm
      ads}^2}+1-\frac{r_0^{d-3}}{r^{d-3}}\right)\,.
\ee
The quantity $r_0$ is related to the mass $M$ of the spacetime,
\be
M=\frac{(d-2)A_{d-2}r_0^{d-3}}{16\pi}\,,
\ee
and the horizon radius $r_+$ is the largest real root of $f(r)=0$.


For circular null geodesics, an exact solution can be found with
\be 
r_c=2^{\frac{1}{3-d}}\left\lbrack(d-1)\left
(1+\frac{r_+^2}{L_{\rm ads}^2}\right
)\right\rbrack^{\frac{1}{d-3}}r_+ \,.
\label{nullcircleAdS}
\ee
Since $V_r''(r_c)=L^2 \left (2d-6\right ) r_c^{-4}$, circular null geodesics
are unstable.  The angular velocity at this radius is given by
\be L_{\rm ads}^2\,\Omega_c^2=1+
\frac{(d-3)}{2}
\left(\frac{2}{d-1}\right)^{\!\!\frac{d-1}{d-3}}
{\left(1+\frac{r_+^2}{L_{\rm ads}^2}\right)^{\!\!-\frac{2}{d-3}}}
\frac{L_{\rm ads}^2}{r_+^2} \,,
\ee
and reduces to $\Omega_c \approx 1/L_{\rm ads}$ for large $(r_+/L_{\rm ads}\gg
1)$ black holes.  The calculation of the instability exponents proceeds
trivially. As in the case of higher dimensional Schwarzschild black holes,
Eq. (\ref{tangh}), we find 
\be
\lambda/\Omega_c=\sqrt{d-3}\,.
\ee
If our main result, Eq.~(\ref{QNMLyapunov}), were valid in asymptotically AdS
spacetimes, in the eikonal limit we would get $\omega_{\rm
  QNM}/\Omega_c=l-i\sqrt{d-3}\,(n+1/2)$. However, according to both analytical
\cite{Festuccia:2008zx} and numerical results
\cite{Horowitz:1999jd,Cardoso:2003cj,Miranda:2005qx,Berti:2003zu}, the
imaginary part of QNM frequencies in this background {\it increases
  monotonically} with increasing $r_+$.  This dependence on $r_+$ cannot be
explained by the circular null geodesic analogy.  In hindsight, this failure
is not too surprising. AdS spacetimes are not globally hyperbolic; boundary
conditions at infinity must be taken into account. On the other hand geodesic
calculations are local, and they carry no information about spatial infinity.
The available analytical and numerical results
\cite{Festuccia:2008zx,Horowitz:1999jd,Cardoso:2003cj,Miranda:2005qx,Berti:2003zu}
indicate that the damping timescale is {\it smaller} than indicated by the
geodesic calculation, Eq.~(\ref{qnadsgeo}). Perhaps the disagreement could be
explained by arguing that null particles reach spatial infinity on timescales
faster than the geodesic timescale, and therefore one would have to correct
for this.




\begin{thebibliography}{99}

\bibitem{Narayan:2005ie}
  R.~Narayan,
  New J.\ Phys.\  {\bf 7}, 199 (2005)
  [arXiv:gr-qc/0506078].

\bibitem{Zhang:1997dy}
 S.~N.~Zhang, W.~Cui and W.~Chen,
 Astrophys.\ J.\  {\bf 482}, L155 (1997)
 [arXiv:astro-ph/9704072].
 
\bibitem{Shapiro:1983du}
 S.~L.~Shapiro and S.~A.~Teukolsky,
 {\it Black holes, white dwarfs, and neutron stars: The physics of compact
 objects} (Wiley, New York, 1983).
 
\bibitem{podurets} M. A. Podurets, Astr. Zh. {\bf 41}, 1090 (1964) (English translation in Sovet Astr.-AJ {\bf 8}, 868 (1965)).
 
\bibitem{amesthorne} W. L. Ames and K. S. Thorne,
Astrophys.\ J.\ {\bf 151}, 659 (1968).  
 
\bibitem{Kokkotas:1999bd}
 K.~D.~Kokkotas and B.~G.~Schmidt,
 Living Rev.\ Rel.\  {\bf 2}, 2 (1999)
 [arXiv:gr-qc/9909058].

\bibitem{Nollert:1999ji}
 H.~P.~Nollert,
 Class.\ Quant.\ Grav.\  {\bf 16}, R159 (1999).


\bibitem{pressringdown} W. H. Press,
Astrophys. J. {\bf 170}, L105 (1971).

\bibitem{goebel} C. J. Goebel,
Astrophys. J. {\bf 172}, L95 (1972).

\bibitem{Ferrari:1984zz} V.~Ferrari and B.~Mashhoon,
Phys.\ Rev.\  D {\bf 30}, 295 (1984).

\bibitem{mashhoon} B. Mashhoon,
Phys. Rev. {\bf D31}, 290 (1985).

\bibitem{Berti:2005eb}
  E.~Berti and K.~D.~Kokkotas,
  Phys.\ Rev.\  D {\bf 71}, 124008 (2005)
  [arXiv:gr-qc/0502065].

\bibitem{Bombelli:1991eg}
 L.~Bombelli and E.~Calzetta,
 Class.\ Quant.\ Grav.\  {\bf 9}, 2573 (1992).
 
\bibitem{Cornish:2003ig}
 N.~J.~Cornish and J.~J.~Levin,
 Class.\ Quant.\ Grav.\  {\bf 20}, 1649 (2003)
 [arXiv:gr-qc/0304056].
 
\bibitem{Pretorius:2007jn}
 F.~Pretorius and D.~Khurana,
 Class.\ Quant.\ Grav.\  {\bf 24}, S83 (2007)
 [arXiv:gr-qc/0702084].
 
\bibitem{PerezGiz:2008yq}
  G.~Perez-Giz and J.~Levin,
  arXiv:0811.3815 [gr-qc].

\bibitem{Levin:2008yp}
  J.~Levin and G.~Perez-Giz,
  arXiv:0811.3814 [gr-qc].

\bibitem{Grossman:2008yk}
  R.~Grossman and J.~Levin,
  arXiv:0811.3798 [gr-qc].

\bibitem{Steklain:2008gz}
  A.~F.~Steklain and P.~S.~Letelier,
  Phys.\ Lett.\  A {\bf 373}, 188 (2009)
  [arXiv:0811.4049 [gr-qc]].


\bibitem{Sperhake:2008ga}
  U.~Sperhake, V.~Cardoso, F.~Pretorius, E.~Berti and J.~A.~Gonzalez,
  Phys.\ Rev.\ Lett.\  {\bf 101}, 161101 (2008)
  [arXiv:0806.1738 [gr-qc]].
 
\bibitem{Shibata:2008rq}
  M.~Shibata, H.~Okawa and T.~Yamamoto,
  Phys.\ Rev.\  D {\bf 78}, 101501 (2008)
  [arXiv:0810.4735 [gr-qc]].

\bibitem{Mashhoon:1983}
B. Mashhoon, in {\em Proceedings of the Third Marcel Grossmann Meeting
on Recent Developments of General Relativity, Shanghai, 1982}, edited
by Hu Ning (North-Holland, Amsterdam, 1983).

\bibitem{Schutz:1985km} B.~F.~Schutz and C.~M.~Will,
Astrophys. J. {\bf 291}, L33 (1985).

\bibitem{Iyer:1986np}
  S.~Iyer and C.~M.~Will,
  Phys.\ Rev.\  D {\bf 35}, 3621 (1987).

\bibitem{Iyer:1986nq}
  S.~Iyer,
  Phys.\ Rev.\  D {\bf 35}, 3632 (1987).

\bibitem{Tangherlini:1963bw}
 F.~R.~Tangherlini,
 Nuovo Cim.\  {\bf 27}, 636 (1963).

\bibitem{Cardoso:2003sw}
 V.~Cardoso and J.~P.~S.~Lemos,
 Phys.\ Rev.\  D {\bf 67}, 084020 (2003)
 [arXiv:gr-qc/0301078].

\bibitem{Myers:1986un}
 R.~C.~Myers and M.~J.~Perry,
 Annals Phys.\  {\bf 172}, 304 (1986).

\bibitem{Emparan:2003sy}
 R.~Emparan and R.~C.~Myers,
 JHEP {\bf 0309}, 025 (2003)
 [arXiv:hep-th/0308056].

\bibitem{Bardeen:1972fi}
 J.~M.~Bardeen, W.~H.~Press and S.~A.~Teukolsky,
 Astrophys.\ J.\  {\bf 178}, 347 (1972).

\bibitem{chaos} J. R. Dorfman, {\it An Introduction to Chaos in Nonequilibrium Statistical Mechanics}
(Cambridge University Press, Cambridge, 1999); H. A. Posh and W. G. Hoover, J. Phys.: Conf. Ser. {\bf 31}, 9 (2006).

\bibitem{skokos} C. Skokos,
%
arXiv:0811.0882 [nlin.CD].

\bibitem{stewart} J. M. Stewart,
Proc. R. Soc. London {\bf A424}, 239 (1989).

\bibitem{MTB} S. Chandrasekhar, {\it The Mathematical Theory of Black Holes},
(Oxford University Press, New York, 1983).

\bibitem{Kodama:2003jz}
  H.~Kodama and A.~Ishibashi,
  Prog.\ Theor.\ Phys.\  {\bf 110}, 701 (2003)
  [arXiv:hep-th/0305147].

\bibitem{Ishibashi:2003ap}
  A.~Ishibashi and H.~Kodama,
  Prog.\ Theor.\ Phys.\  {\bf 110}, 901 (2003)
  [arXiv:hep-th/0305185].

\bibitem{Kodama:2003kk}
  H.~Kodama and A.~Ishibashi,
  Prog.\ Theor.\ Phys.\  {\bf 111}, 29 (2004)
  [arXiv:hep-th/0308128].

%

\bibitem{Rosa:2008dh}
  V.~M.~Rosa and P.~S.~Letelier,
  Phys.\ Rev.\  D {\bf 78}, 084038 (2008)
  [arXiv:0810.1177 [gr-qc]].

\bibitem{merrick} C. Merrick and F. Pretorius, unpublished (2007).

\bibitem{barreto} A. S. Barreto and M. Zworski,
Math. Res. Lett. {\bf 4}, 103 (1997).

\bibitem{Konoplya:2003ii}
 R.~A.~Konoplya,
 Phys.\ Rev.\  D {\bf 68}, 024018 (2003)
 [arXiv:gr-qc/0303052].

\bibitem{Berti:2003si}
 E.~Berti, M.~Cavaglia and L.~Gualtieri,
 Phys.\ Rev.\  D {\bf 69}, 124011 (2004)
 [arXiv:hep-th/0309203].

\bibitem{Cardoso:2004cj}
 V.~Cardoso, G.~Siopsis and S.~Yoshida,
 Phys.\ Rev.\  D {\bf 71}, 024019 (2005)
 [arXiv:hep-th/0412138].

\bibitem{Ida:2002ez}
 D.~Ida, K.~y.~Oda and S.~C.~Park,
 Phys.\ Rev.\  D {\bf 67}, 064025 (2003)
 [Erratum-ibid.\  D {\bf 69}, 049901 (2004)]
 [arXiv:hep-th/0212108].

\bibitem{Ida:2006tf}
 D.~Ida, K.~y.~Oda and S.~C.~Park,
 Phys.\ Rev.\  D {\bf 73}, 124022 (2006)
 [arXiv:hep-th/0602188].

\bibitem{Ida:2005ax}
 D.~Ida, K.~y.~Oda and S.~C.~Park,
 Phys.\ Rev.\  D {\bf 71}, 124039 (2005)
 [arXiv:hep-th/0503052].

\bibitem{Cardoso:2005vk}
 V.~Cardoso and S.~Yoshida,
 JHEP {\bf 0507}, 009 (2005)
 [arXiv:hep-th/0502206].

\bibitem{Frolov:2003en}
 V.~P.~Frolov and D.~Stojkovic,
 Phys.\ Rev.\  D {\bf 68}, 064011 (2003)
 [arXiv:gr-qc/0301016].

\bibitem{Berti:2003yr}
  E.~Berti, K.~D.~Kokkotas and E.~Papantonopoulos,
  Phys.\ Rev.\  D {\bf 68}, 064020 (2003)
  [arXiv:gr-qc/0306106].

\bibitem{Berti:2005ys}
 E.~Berti, V.~Cardoso and C.~M.~Will,
 Phys.\ Rev.\  D {\bf 73}, 064030 (2006)
 [arXiv:gr-qc/0512160].

\bibitem{Berti:2007fi}
  E.~Berti, V.~Cardoso, J.~A.~Gonzalez, U.~Sperhake, M.~Hannam, S.~Husa and B.~Bruegmann,
  Phys.\ Rev.\  D {\bf 76}, 064034 (2007)
  [arXiv:gr-qc/0703053].

\bibitem{Hanna:2008um}
  C.~Hanna, M.~Megevand, E.~Ochsner and C.~Palenzuela,
  arXiv:0801.4297 [gr-qc].

\bibitem{Baker:2008mj}
 J.~G.~Baker, W.~D.~Boggs, J.~Centrella, B.~J.~Kelly, S.~T.~McWilliams and J.~R.~van Meter,
 Phys.\ Rev.\  D {\bf 78}, 044046 (2008)
 [arXiv:0805.1428 [gr-qc]].

\bibitem{Fidkowski:2003nf}
 L.~Fidkowski, V.~Hubeny, M.~Kleban and S.~Shenker,
 JHEP {\bf 0402}, 014 (2004)
 [arXiv:hep-th/0306170].

\bibitem{Amado:2008hw}
  I.~Amado and C.~Hoyos-Badajoz,
  JHEP {\bf 0809}, 118 (2008)
  [arXiv:0807.2337 [hep-th]].
  
\bibitem{Festuccia:2008zx}
 G.~Festuccia and H.~Liu,
 arXiv:0811.1033 [gr-qc].

\bibitem{Banados:1992wn}
  M.~Banados, C.~Teitelboim and J.~Zanelli,
  Phys.\ Rev.\ Lett.\  {\bf 69}, 1849 (1992)
  [arXiv:hep-th/9204099].

\bibitem{Cardoso:2001hn}
 V.~Cardoso and J.~P.~S.~Lemos,
 Phys.\ Rev.\  D {\bf 63}, 124015 (2001)
 [arXiv:gr-qc/0101052].

\bibitem{Horowitz:1999jd}
 G.~T.~Horowitz and V.~E.~Hubeny,
 Phys.\ Rev.\  D {\bf 62}, 024027 (2000)
 [arXiv:hep-th/9909056].


\bibitem{Cardoso:2003cj}
 V.~Cardoso, R.~Konoplya and J.~P.~S.~Lemos,
 Phys.\ Rev.\  D {\bf 68}, 044024 (2003)
 [arXiv:gr-qc/0305037].


\bibitem{Miranda:2005qx}
 A.~S.~Miranda and V.~T.~Zanchin,
 Phys.\ Rev.\  D {\bf 73}, 064034 (2006)
 [arXiv:gr-qc/0510066].

\bibitem{Berti:2003zu}
  E.~Berti and K.~D.~Kokkotas,
  Phys.\ Rev.\  D {\bf 68}, 044027 (2003)
  [arXiv:hep-th/0303029].

\end{thebibliography}
\end{document}